\documentstyle[aps,pre]{revtex}
\begin{document}
\draft
\twocolumn

\title{State Selection in Accelerated Systems}

\author{Martin B. Tarlie$^1$  and K. R. Elder$^{2}$
}

\address{$^1$James Franck Institute,
University of Chicago, 5640 South Ellis Avenue, 
Chicago, IL 60637} 
\address{$^2$Department of Physics, Oakland University, Rochester, MI,
48309-4487}

\date{\today}

\maketitle

\begin{abstract}

The problem of state selection when multiple metastable states compete
for occupation is considered for systems that are accelerated far 
from equilibrium. The dynamics of the supercurrent in a narrow 
superconducting ring under the influence of an external electric field
is used to illustrate the general phenomenology.  

\end{abstract}


%\begin{multicols}{2}

Many systems when driven far from equilibrium encounter instabilities 
that lead towards new states or phases.  Frequently there exist multiple 
states that can be selected following the onset of the instability.  The 
determination of the particular state that is selected is a complex problem 
of fundamental interest in a wide variety of fields\cite{ch93,ksz88}.
In addition to the complexity associated with the presence of 
multiple competing states, if the system is accelerated so that the 
environment evolves in time, then the state that is selected can 
depend in an important way on the driving force. The focus of this 
paper is on state-selection in ``accelerated'' systems when multiple 
metastable states compete for occupation.

For the purposes of this paper an accelerated system is defined 
as one for which a control parameter is varied in time so that the 
system gradually progresses from stable to unstable regimes.  For 
example, in a narrow superconducting 
ring\cite{la67,mh70,t95,t80} under the influence of 
a constant electromotive force, the superconducting electrons 
are accelerated by the electric field and the supercurrent increases with 
time until the critical current is reached and the system becomes
(Eckhaus) unstable.  Similar behavior could occur in direction
solidification\cite{ms64,l80}
if the solidification cell is accelerated slowly, rather than pulled at
a constant velocity, through a temperature gradient until
the (Mullins-Sekerka) instability is encountered and the
liquid/solid interface becomes unstable.
In each of these scenarios the systems become unstable with respect 
to fluctuations of certain wavelengths that lie within a band.
When the size of the system is comparable to the length scales 
associated with the wavevectors in the unstable band, the system is 
described as mesoscopic and the number of accessible states is finite. 
As illustrated in an extensive review by Cross and
Hohenberg\cite{ch93}, instabilities that result in this type 
of mesoscopic behavior are extremely common, occurring in many diverse 
fields, such as fluid dynamics, chemical reactions, material science 
and biology.

The focus of this paper is on state selection in mesoscopic
accelerated systems immediately following the onset of the
instability.  The combination of the driving
force that accelerates the system, and the mesoscopic system
size that allows for multiple, isolated metastable states,
leads to novel and interesting selection rules.
It will be shown that the rate at which the system is driven 
through the instability plays a prominent role in determining 
the probability that a particular metastable state is 
selected.
As the decay is from states of marginal stability, the selection is
also influenced by the noise in the system.
The dependences on both acceleration rate and noise strength
are considered.

For illustrative purposes consider the situation in which an infinitely
long solenoid, carrying a current that increases linearly with time, 
passes through the center of a narrow superconducting ring of
cross-sectional area $A$ and 
circumference $L\equiv \xi(T) \ell$, where $\xi(T)$ is the 
temperature dependent
correlation length. By Faraday's law of induction, 
a constant electromotive force (emf) $V$ is induced in the 
superconductor, thereby accelerating the superconducting electrons.
The dynamics of the 
(dimensionless) superconducting order parameter $\psi(x,t)$, 
where $x$ is the longitudinal spatial coordinate and $t$ is time, 
is described by the stochastic time-dependent Ginzburg-Landau 
equation\cite{mh70,t95}:
\begin{equation}
\label{eq:stdgl}
\partial_{t}\psi =  \partial_{x}^{2}\psi + \psi
-\psi |\psi |^{2} + i\ell^{-1}x\omega\psi + \eta
\end{equation} 
where $\omega\equiv\tau_{GL}(2eV/\hbar)$ is a dimensionless 
measure of the strength of the induced emf.  
Throughout this paper the regime where $\omega \ll 1$ is
considered.  In Eq.~(\ref{eq:stdgl}) $\tau_{GL}$, is the 
Ginzburg-Landau relaxation time,
and $\psi$ satisfies the twisted-periodic boundary
condition $\psi (\ell +x,t)=\exp (i\omega t)\psi (x,t)$.
This equation ignores the influence of the normal current and
is valid for a voltage driven system 
in the limit of low normal state resistivity.
The variable $\eta$ is a Gaussian random variable, 
with expectation values $\langle\eta (x,t)\rangle =0$, and 
$\langle\eta (x,t)\eta^{*}(x',t')\rangle =
2D\delta (x-x')\delta (t-t')$,
where $D$ is determined by the fluctuation-dissipation 
theorem\cite{dval}.

For $\omega\! \ll \! 1$, the relevant current-carrying states of the 
superconductor are uniformly twisted plane wave solutions
given by $\bar\psi=\sqrt{1-q^2}\exp (iqx)$, where $q=mK+\omega t/\ell$, 
and $K\equiv 2\pi/\ell$. 
The dimensionless current density $J$ of these
states is given by $J=(\bar\psi^*\partial_x\bar\psi
- \bar\psi\partial_x\bar\psi^*)/2i = q(1-q^2)$.  Thus the effect of 
the induced emf (which increases $q$ linearly with time) is to wind 
the order parameter, or equivalently, to accelerate the superconducting 
electrons.  However, this acceleration
cannot continue indefinitely because $J$ is a nonmonotonic function
of $q$, and hence time, achieving a maximum value of $J_{c}=2/\sqrt{27}$ 
at $q=q_{c}=1/\sqrt{3}$. 
This saturation of the current at the critical
current $J_{c}$, coincides with the loss of stability of states 
$\bar\psi$ at $q=q_{c}$. In other words, for $q>q_{c}$, 
$d^2F(q)/dq^2<0$, where $F(q)\equiv F_{GL}[\bar\psi]$ is the Ginzburg-Landau
free energy of states $\bar\psi$.

To understand the Eckhaus instability for finite size systems it
is necessary to perform a linear stability analysis about the
state $\bar\psi$, as the previous analysis only applies 
when $\ell =\infty$.
Standard linear stability analysis gives one potentially positive  
eigenvalue that takes the form\cite{kz85}
\begin{equation}
\lambda_n(q) =
-1+q^{2}-k_n^{2}+
\sqrt{(1-q^2)^2+4q^2k_n^2}\,\,.
\label{eq:lambda}
\end{equation}
The eigenvector associated with this eigenvalue is a linear 
combination of Fourier modes with wavevector $q \pm k_n$ and 
amplitude $A_n$, where $k_n=nK$. 
The interesting feature of this eigenvalue 
is that it can become positive when $q > \kappa_{1}>q_{c}$,  
where $\kappa_m\equiv \frac{1}{\sqrt{3}}[1+m^2K^2 /2]^{1/2}$. 
Thus, for finite size systems the instability is pushed to wavevectors
greater than $q_{c}$ by an amount that depends on $\ell$\cite{kz85}.
In particular, 
for $\kappa_m > q > \kappa_1$, $\lambda_n$ is positive 
for all values of $k_n < mK$ \cite{catalysts}.   The dependence 
of $\lambda$ on $k_n$ is shown in the inset of  
Fig.~\ref{fig:lindisp} for several values of $q$.

The growth of a single Fourier mode 
(with amplitude $A_{n}$) 
of wavevector $q-k_n$, and simultaneous decay of $A_0$,
corresponds to a decrease of the winding 
number $W=(2\pi)^{-1}\int_{0}^{\ell}d\phi (x)/dx$, where $\phi$ is
the phase of $\psi$, by an amount $n$.  This phenomenon is known 
as a ``phase-slip'' as the total phase of the order parameter
changes by an integral multiple of $2\pi$. Physically, the 
supercurrent decreases by a discrete amount when a phase-slip 
occurs. Phase-slip processes can also occur via thermal activation 
over an energy barrier 
and this process has received significant attention over the 
years\cite{la67,mh70,t80,tsg94}.  For $D=10^{-3}$, as long as 
$\omega\agt 10^{-24}$, the probability of a thermally 
activated phase slip occurring is exceedingly small\cite{t95}. 
Thus, unless the temperature is very close to the superconducting 
transition temperature $T_{c}$, where $D$ is large, the system 
will almost always be driven to the Eckhaus instability before a 
thermally activated phase-slip can occur. Therefore, the transitions 
that are of concern in this work involve the decay from an 
unstable state, in contrast to previous 
work\cite{la67,mh70,t95,t80} where the focus was on the decay 
from a metastable state. 

When $\omega > 0$, the system is driven to the point of instability 
as the eigenvalues of each Fourier mode eventually become positive. 
As illustrated in Fig.~\ref{fig:lindisp},
the $n=1$ mode becomes unstable first, 
then the $n=2$ mode becomes unstable, and so on.
This implies that the system first becomes unstable with respect to  
single phase-slip processes, then double phase-slip processes, etc. 
If $\omega$ is large enough, the $n=1$ mode might 
not have time to grow to dominance by the time the $n=2$ mode becomes
unstable. This suggests that for small $\omega$, single phase-slip 
processes should dominate the dynamics, but as $\omega$ is 
increased there is a crossover to a regime in which double phase-slip 
processes dominate. 
As $\omega$ is increased further,
double phase-slip processes should give way to triple phase-slip 
processes, and so on.

The generic features displayed in Fig.~\ref{fig:lindisp} are common 
to many systems and come under the general classification scheme 
of Cross and Hohenberg\cite{ch93} as type ${\rm II}_{\rm s}$. Thus, the 
dynamic competition between unstable modes discussed above
is a phenomena that has relevance to many systems. 
The precise determination of which of the modes will initially 
be selected following the onset of the instability is a complex question 
that depends on the details of the individual system.  
For concreteness, the general problem of the dependence of the selected
state as a function of the driving force will be addressed for the 
superconductor. In particular, the relative probability that a 
given state is selected will be determined by extensive computer simulations.
Second, it will be shown 
that the qualitative features of these results can be understood by 
an analysis that is based on the properties of the growth rates
$\lambda$.

To evaluate the probability of the occurrence of 
a given phase slip as a function of $\omega$, 
Eq. (\ref{eq:stdgl}) was numerically integrated in time  
for a noise strength of $D=10^{-3}$ and a length corresponding 
to $n_{\ell} \equiv \ell q_{c}/(2\pi)=5$\cite{ref:nl}.
In Fig.~\ref{fig:pvsw}a, the probability of a type-$n$ phase slip 
is plotted as a function of $\omega$. As expected, for small $\omega$, 
single phase-slips dominate. As $\omega$ increases further there is a 
crossover to a regime in which double phase-slips dominate. Further 
increase of $\omega$ results in a subsequent crossover to a regime in
which triple phase-slips dominate, and so on. 

An example of the dynamics that lead to such results is shown in 
Fig.~\ref{fig:dynam}. In this figure, the winding number and 
current are plotted as a function of time for $\omega =5\times 10^{-4}$.
This value of $\omega$ is in the crossover region between the single 
and double phase-slip dominated regimes. Clearly evident in this figure
are the single and double phase slips in which $W$ changes by one or
two, respectively. Also seen in Fig.~\ref{fig:dynam}b are the 
discrete jumps of the supercurrent.

As described earlier, the essential features shown in 
Fig.~\ref{fig:pvsw}a can be understood using the properties of 
the growth rates $\lambda_n$. This idea can be made more 
concrete in the following way. Ignoring the nonlinear interactions
between the different modes, the expectation of 
$|A_{n}|^2$ is given by\cite{t95}
\begin{equation}
\langle |A_{n}(t)|^2 \rangle  = {2D\over{\ell}}{e}^{2\sigma_{n}(t)}
\int_{0}^{t}dt' \,{e}^{-2\sigma_{n}(t')},
\label{eq:a2ou}
\end{equation}
where $\sigma_{n}(t)\equiv\int_{0}^{t}dt'\,
\lambda_{n}\left( q(t')\right) $, and
angular brackets denote a noise average.
After the onset of the instability the system evolves 
towards the fixed points $\bar\psi_{n}=\bar A_{n}\exp (i(q-nK)x)$, 
where $\bar{A}_n\!=\!\sqrt{1-(q-nK)^2}$.
Eq.~(\ref{eq:a2ou}) describes the initial evolution of the system
after the Eckhaus boundary has been reached.
In this non-interacting picture 
each amplitude (measured in units of $\bar A_{n}$)
can be thought of as an orthogonal coordinate 
in an $n_{\ell}$-dimensional space. Thus, the natural measure of distance 
from the origin ($A_{n}=0$) in this space  
is $\sum_{n=1}^{n_{\ell}}\langle |A_n(t)|^2\rangle /\bar{A}^2_n$.  After 
onset of the Eckhaus instability, this sum increases rapidly 
and reaches unity at a time $t^*$. Assuming  
that at $t=t^*$ a phase-slip has occured with probability one it 
is natural to interpret
$\langle |A_{n}(t^{*})|^2\rangle/\bar{A}^2_n$ as
the relative probability of the occurrence of a type-$n$ phase-slip.  
The probabilities calculated using this procedure are 
shown in Fig.~\ref{fig:pvsw}b.

It is clear from Fig.~\ref{fig:pvsw} that the preceding analysis
provides a qualitatively accurate description of the state-selection
probabilities, and their dependence on the driving force $\omega$. 
Most notably, the values of $\omega$ at the peak 
positions agree very well with the numerical results.
Nevertheless it is important to point out that the preceding 
analysis is only a plausible argument and is not 
systematic.  A quantitative description must include the 
subtle non-linear interactions that are an important element 
in determining state selection.  Even at the present 
level of ignorance, however, the analysis presented here 
provides a qualitatively useful description of the state-selection 
probabilities, and their dependence on the driving force.  

The growth rates $\lambda$ are an extremely important factor in 
determining the state selection probabilities. The preceding 
analysis accounts for these growth rates and therefore provides
a qualitatively accurate description.  The analysis also 
provides predictions for the dependence of $P_n$ on the noise 
strength $D$, which may be more convenient to vary in some 
experiments. Plotted in Fig.~\ref{fig:pvsd}a are the probabilities 
of a type-$n$
phase slip as a function of $D$, for a fixed value of $\omega$, 
obtained from a numerical simulation of Eq.~(\ref{eq:stdgl}). 
In Fig.~\ref{fig:pvsd}b, the corresponding $P_{n}$'s obtained from 
the growth-rate analysis are plotted for comparison.  Once again, 
it is seen that the simple analysis provides an accurate qualitative 
picture. For the smallest values of $D$ considered 
triple phase-slip processes dominate.  
This is because the time required for a given mode to grow 
to saturation diverges logarithmically as $D\rightarrow 0$. Consequently, 
if $D$ is very small, the mode amplitudes $A_1$ and $A_2$, for example, may 
still be very small by the time the growth rate of $A_3$ is 
significantly larger than the growth rates for $A_1$ or $A_2$.

One of the most interesting aspects of the phenomena 
exposed here is that the selection rules depend on both the intrinsic
properties of the system and the external parameters.
To understand this connection more deeply, it is instructive to 
consider the characteristic growth times for individual modes.
Typically, the characteristic time associated with the initial 
growth of an unstable mode is taken to be the inverse of the 
growth rate. However, for accelerated systems the growth rate $\lambda$ 
starts out negative and passes through zero. Thus,
$|\lambda^{-1}|$ is not a relevant quantity as it diverges at the 
instability. To determine the characteristic time, 
consider Eq.~(\ref{eq:a2ou}) for
$\langle |A_{n}(t)|^{2}\rangle$. The quantity $\sigma (t)$ achieves 
a local minimum at $t=t_{n}\equiv \ell\kappa_{n}/\omega$ so that a 
second order expansion about $t_{n}$ yields 
$\sigma_{n}(t)\approx \sigma_{n}(t_{n})+
\frac{1}{2}\frac{\lambda_{n}^{\prime}\omega}{\ell}(t-t_{n})^{2},$
where $\lambda_{n}^{\prime}\equiv 
\partial\lambda_{n} /\partial q|_{q=\kappa_n}$. 
Inserting this expansion into Eq.~(\ref{eq:a2ou}) and assuming that 
$\omega\ll 1$ gives    
\begin{equation}
\langle |A_{n}(t)|^{2}\rangle =
2D\tau_{n}\ell^{-1}\exp\left( z_{n}^{2}(t)\right)
\left[ {\rm erf}\left( z_{n}(t)\right) +1\right] ,
\label{eq:Aapp}
\end{equation}
where $z_{n}(t)=(t-t_{n})/\tau_{n}$ and 
$\tau_{n}=\sqrt{\ell/\lambda_{n}^{\prime}\omega}$. 
The quantity $\tau_{n}$ is the characteristic time for the 
growth of mode-$n$, and is interesting because it depends 
on the geometric mean of $\lambda_{n}^{\prime}$ and $\omega$. 
Thus, the time scale $\tau_{n}$ embodies in a natural way the 
importance of the combination of the intrinsic 
dynamics ($\lambda_{n}^{\prime}$) and the external driving force ($\omega$). 

In summary, the problem of state selection in accelerated systems has been 
shown to contain unique and rich phenomenology. Although the focus 
of this work has been on the dynamics of quasi-one-dimensional superconducting
rings, the essential features of this particular system are generic and 
should be observable in a diverse array of experimentally realizable 
situations. Despite the success of the linear analysis,
it is clear that new methods must be developed to 
explore this complex and important area of research in nonequilibrium 
statistical mechanics. Recent work\cite{tm97} on state selection 
in non-accelerated marginally stable systems
suggests a possible systematic framework that could be extended to
address the phenomena considered here. 

We thank Paul Goldbart for suggesting the problem of nonequilibrium 
superconductivity, and Paul Goldbart and Alan McKane for useful discussions. 
This work was supported in part by the MRSEC program of the NSF (DMR-9400379)
(MBT), Research Corporation Grant
\#CC4181 (KRE), and grant NSF-DMR-8920538 administered through the 
U. of Illinois Materials Research Laboratory (MBT, KRE).

\begin{figure}[btp]
\narrowtext
\caption{$\lambda$ as a function of $q=\omega t/\ell$ for 
$k=K,2K,3K$. Inset: $\lambda$ as a function of $k$ for two values
of $q>\kappa_{1}$ such that the upper curve corresponds to the 
larger value of $q$. }
\label{fig:lindisp}
\end{figure}

\begin{figure}[btp]
\narrowtext
\caption{State selection probabilities as a function of the
driving force $\omega$.  
Open squares, solid squares, open circles, solid circles and 
open triangles correspond to the probabilities 
$P_1, P_2, P_3, P_4$ and $P_5$, respectively. 
Results of the numerical integration of Eq.~(\ref{eq:stdgl}) 
and those of the linear analysis described in the text are shown in 
Figs.~(a) and (b), respectively.}
\label{fig:pvsw}
\end{figure}

\begin{figure}[btp]
\narrowtext
\caption{Dynamics of winding number (a) and supercurrent 
(b), for $\omega =5\times 10^{-4}$ and $D=10^{-3}$. }
\label{fig:dynam}
\end{figure}

\begin{figure}[btp]
\narrowtext
\caption{State selection probabilities as a function of the
noise strength $D$. The symbols in this figure are identical 
to those in Fig.~\ref{fig:pvsw}. 
Results of the numerical integration of Eq.~(\ref{eq:stdgl}) 
and those of the linear analysis described in the text are shown in 
Figs.~(a) and (b), respectively.}
\label{fig:pvsd}
\end{figure}

%\end{multicols}


\begin{thebibliography}{99}
\bibitem{catalysts} Because all modes with $q>\kappa_1$ are 
unstable, the modes of wavevector $q+k$,
which grow initially according to linear analysis, will be damped by the
nonlinear terms. Therefore, these modes act as catalysts in that they 
play an active role in the growth of the $q-k$ modes, but they
themselves cannot compete for occupation as they are unstable.
\bibitem{ref:nl} The parameter $n_{\ell}$ gives the winding number 
of $\bar\psi$ when $q=q_c$ and is therefore a 
useful way to parameterize the circumference of the wire.
\end{thebibliography}
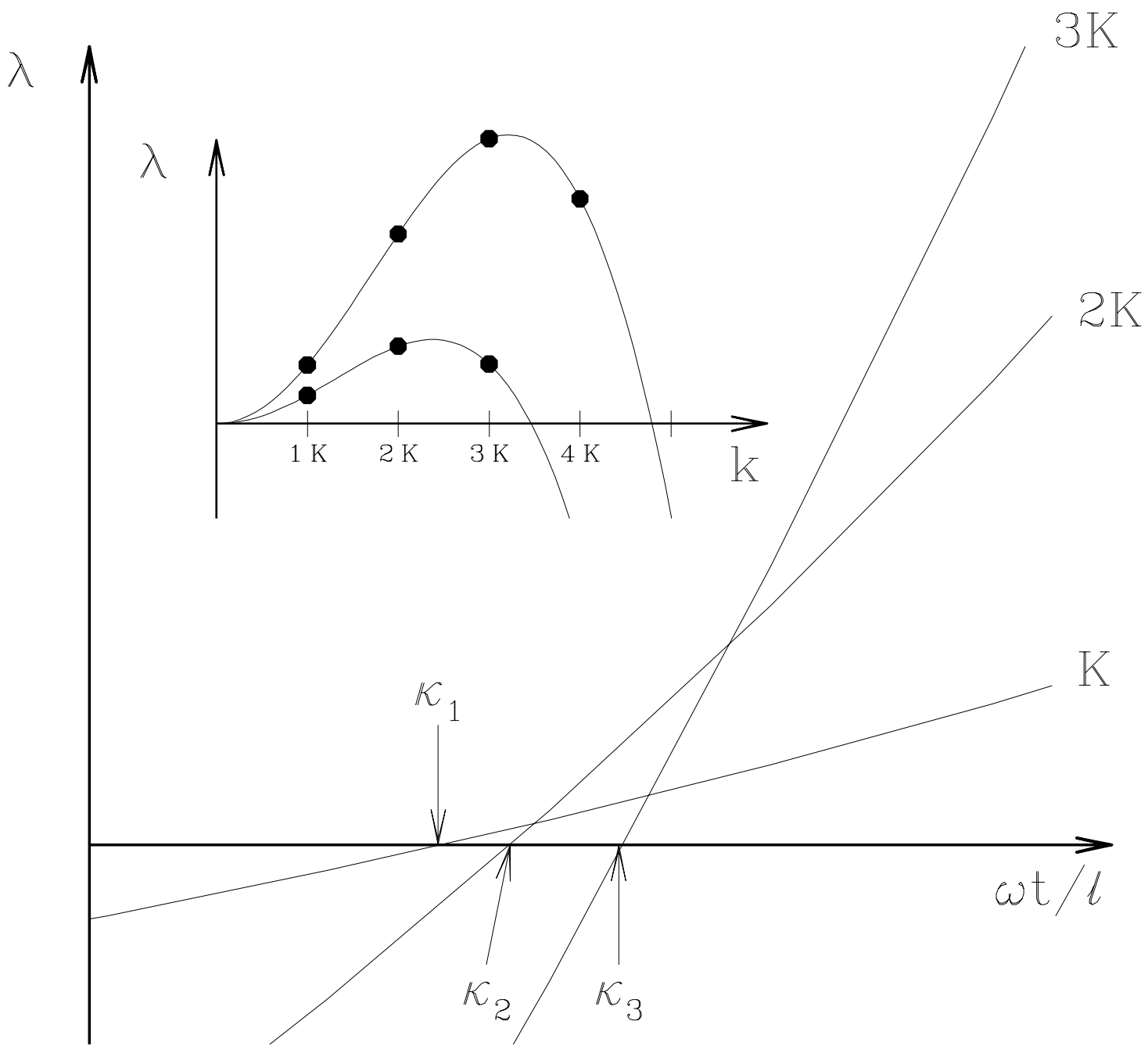
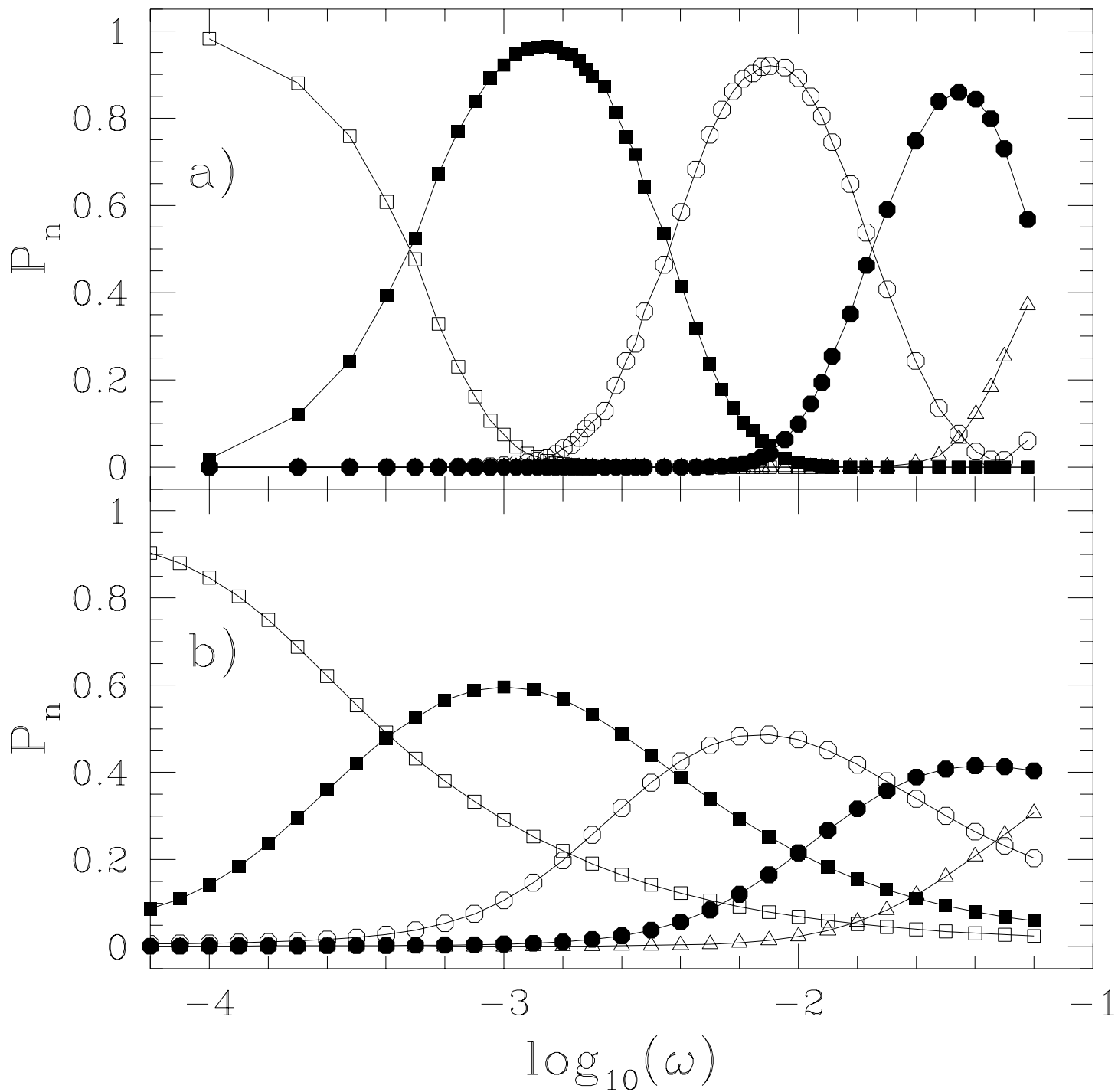
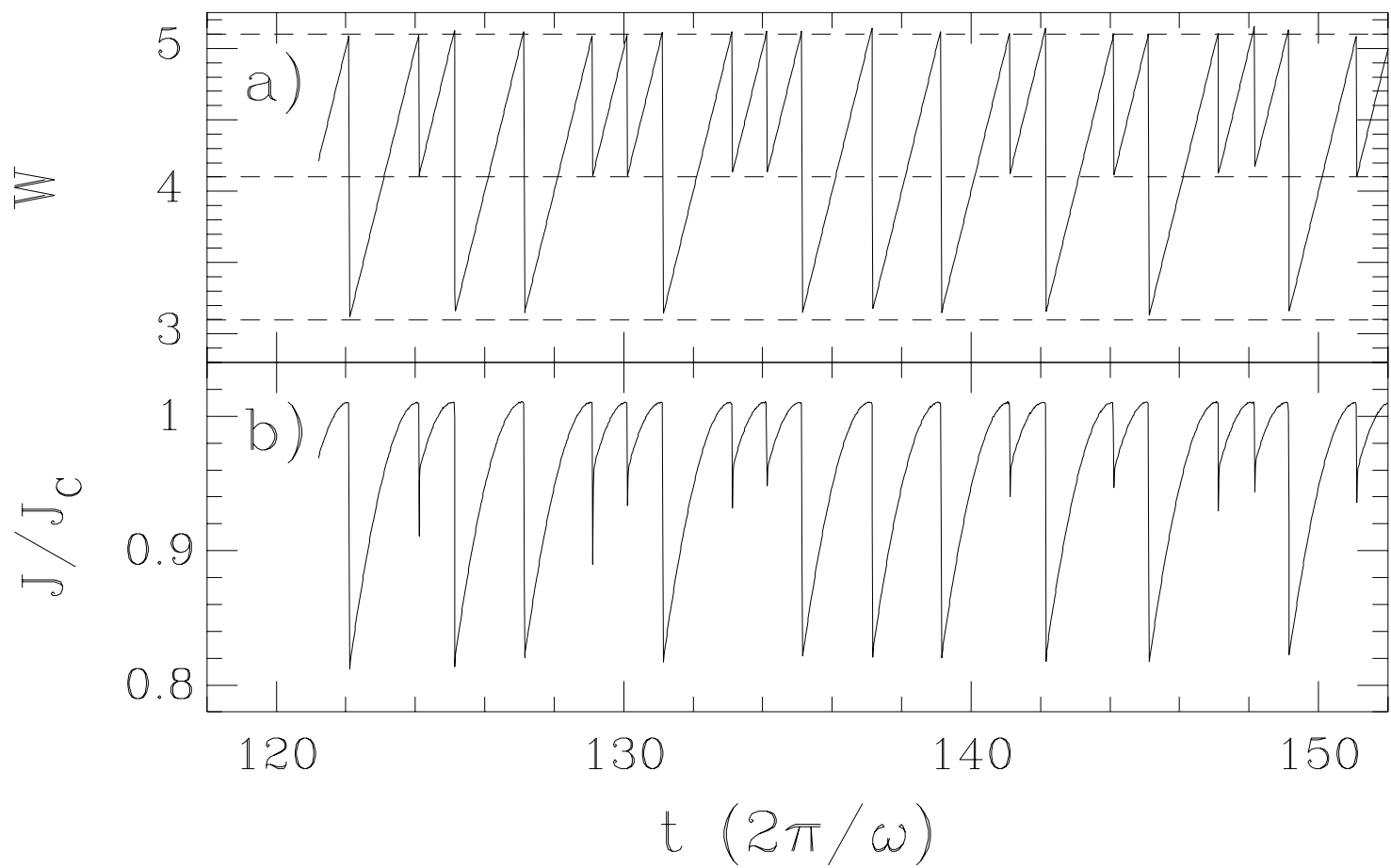
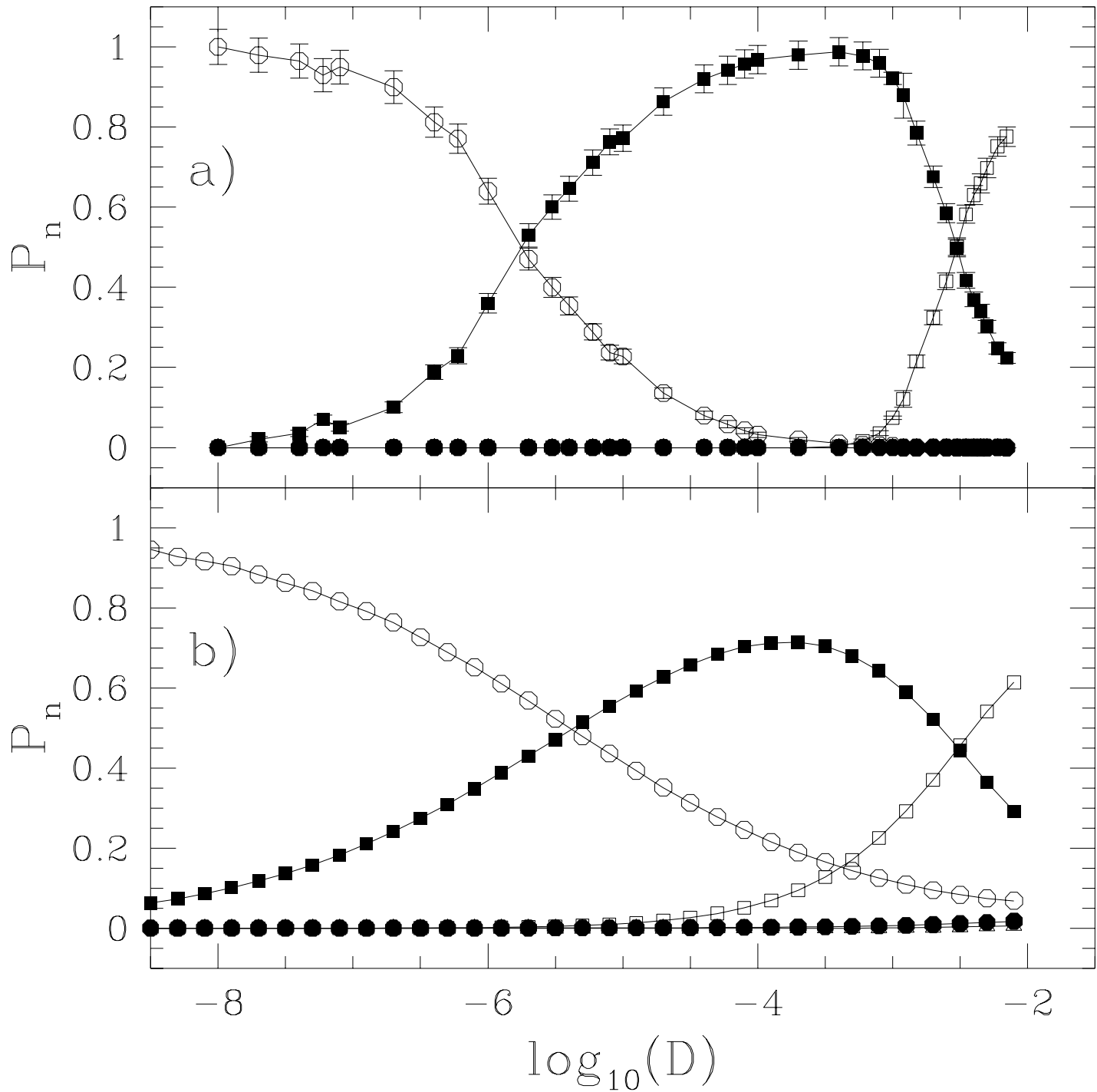
\end{document}